\documentclass[sigconf]{acmart}
\acmConference[EASE 2025]{The 29th International Conference on Evaluation and Assessment in Software Engineering}{17–20 June, 2025}{Istanbul, Türkiye}

\AtBeginDocument{%
  }
\setcopyright{acmlicensed}
\copyrightyear{2025}
\acmYear{2025}
\acmDOI{XXXXXXX.XXXXXXX}
\acmISBN{978-1-4503-XXXX-X/2018/06}
\usepackage{amsmath,amsfonts}
\usepackage{algorithmic}
\usepackage{booktabs}
\usepackage{hyperref}
\usepackage{graphicx}
\usepackage{textcomp}
\usepackage{natbib}
\usepackage{colortbl}
\usepackage[most]{tcolorbox}
\definecolor{main}{HTML}{757677}  
\definecolor{sub}{HTML}{DEE0E3}
\usepackage{mathtools}
\usepackage{hyperref}
\usepackage{multirow}
\usepackage{framed}
\usepackage{xargs}
 
\usepackage[flushleft]{threeparttable}
\usepackage{tabularx}
\usepackage{todonotes}

\colorlet{shadecolor}{gray!10}
\colorlet{framecolor}{black}
{\endMakeFramed}

\newtcolorbox{boxH}{
    colback = sub, 
    colframe = main, 
    boxrule = 0pt, 
    leftrule = 6pt 
}
\begin{document}

\title{Understanding the Impact of Domain Term Explanation on Duplicate Bug Report Detection}

 \author{Usmi Mukherjee}
 \affiliation{
 \institution{Dalhousie University,}
 \city{Halifax}
\state{Nova Scotia}
  \country{Canada}
 }
 \email{usmi.mukherjee@dal.ca}

\author{Mohammad Masudur Rahman}
\affiliation{
\institution{Dalhousie University,}
\city{Halifax}
\state{Nova Scotia}
\country{Canada}
}
\email{masud.rahman@dal.ca}

\begin{abstract}
Duplicate bug reports make up 42\% of all reports in bug tracking systems (e.g., Bugzilla), causing significant maintenance overhead. Hence, detecting and resolving duplicate bug reports is essential for effective issue management. Traditional techniques often focus on detecting textually similar duplicates. However, existing literature has shown that up to 23\% of the duplicate bug reports are textually dissimilar. Moreover, about  78\% of bug reports in open-source projects are very short (e.g., less than 100 words) often containing domain-specific terms or jargon, making the detection of their duplicate bug reports difficult. In this paper, we conduct a large-scale empirical study to investigate whether and how enrichment of bug reports with the explanations of their domain terms or jargon can help improve the detection of duplicate bug reports. We use  92,854 bug reports from three open-source systems, replicate seven existing baseline techniques for duplicate bug report detection, and answer two research questions in this work. We found significant performance gains in the existing techniques when explanations of domain-specific terms or jargon were leveraged to enrich the bug reports. Our findings also suggest that enriching bug reports with such explanations can significantly improve the detection of duplicate bug reports that are textually dissimilar.

\end{abstract}

\maketitle

\section{Introduction} \label{Intro}

\looseness=-1
Software bugs are human-made mistakes that prevent a software system from operating as expected. According to existing studies~\cite{britton2013reversible, zou2018practitioners}, software bugs cause the global economy to suffer enormously and lose billions of dollars every year. Bug finding and corrections also take up $\approx$50\% of a developer's programming time. Bug resolution is, therefore, one of the most challenging tasks in software maintenance~\cite{zou2018practitioners}.

\looseness=-1
Hundreds of software bugs are submitted as \textit{bug reports} to bug-tracking systems such as BugZilla, GitHub and JIRA~\cite{anvik2006should}. Bug tracking systems often receive multiple reports on the same issue (a.k.a., duplicate bug reports) since the users submit their reports independently and asynchronously. Existing studies indicate that up to 42\% of bug reports in a bug tracking system could be duplicates~\cite{zou2018practitioners}, posing significant overhead during software maintenance and evolution~\cite{jalbert2008automated}.

\looseness=-1
Once a software bug is reported, developers need to first understand its root cause and symptoms before they mark the report as a duplicate of previous ones or come up with a solution~\cite{bohme2017directed}. However, they experience several challenges. One major challenge towards bug understanding is missing information in bug reports. Recent studies suggest that information in the majority of bug reports is incomplete and inaccurate~\cite{davies2014s,bugde2008global}. Zhang et al.~\cite{zhang2017bug} found that up to 78\% of bug reports from four open-source projects (e.g., Eclipse, Mozilla, Firefox, GCC) contain less than 100 words each (a.k.a., short bug reports). These short bug reports, on average, took 121 extra days to get resolved by the developers when compared to the well-written bug reports~\cite{zhang2017bug}.  

\looseness=-1
Another significant challenge in understanding bug reports is the prevalent use of domain-specific terms and technical jargon~\cite{zhang2017bug}. These specialized terminologies not only make the bug reports harder to comprehend but also pose significant challenges when detecting their duplicates~\cite{wu2023intelligent}. Different reporters have different levels of technical expertise and thus might use varying amounts of domain terminologies or jargon to describe the same issue. 

\looseness=-1
Another challenge stems from textually dissimilar duplicate bug reports, which make up 19\%--23\% of all duplicate reports~\cite{jahan2023towards}. Existing techniques~\cite{runeson2007detection,wang2008approach,sureka2010detecting,yang2012duplication, aggarwal2017detecting, alipour2013contextual, gopalan2014duplicate, tian2012improved, sun2010discriminative, klein2014new, he2020duplicate,budhiraja2018dwen} are mostly designed for textually similar duplicates. They employ various natural language processing and Information Retrieval (IR) methods, which warrant textual overlap between bug reports for duplicate detection. Thus, these techniques might struggle to detect those duplicate bug reports that are textually dissimilar, rendering them less effective. 

\looseness=-1
Given these challenges above, the prevalence of domain-specific terms in bug reports could be crucial for their understanding or duplicate report detection. These terms carry the core technical aspect or context of an encountered bug, making them potentially valuable indicators of duplicates even when other textual content differs or is missing.
A recent technique~\cite{wu2023intelligent} extracts domain terms from bug reports and leverage IR and deep learning to detect duplicate bug reports. However, it requires both bug reports to contain similar domain terminologies or jargon, which might be less likely. Developers and bug reporters might use different terminologies to express the same issue~\cite{furnas1987vocabulary}. Moreover, the varying technical expertise of reporters could lead to inconsistent use of domain terms, potentially affecting duplicate detection performance. However, despite all these challenges and potential, the impacts of domain-specific terms on the existing techniques for duplicate detection have not been well understood to date, indicating a gap in the literature. Our research addresses this critical knowledge gap.


\looseness=-1
In this paper, we conduct a large-scale empirical study to investigate the impacts of domain-specific terms and how their explanations can affect duplicate bug report detection. Our methodology involves four key steps. First, we target a widely used programming language, Java, and construct comprehensive vocabularies for Java by collecting its domain-specific terms and their explanations from StackOverflow discussions, technical glossaries, and API documentation. Second, we fine-tune an encoder-decoder-based transformer model (e.g., T5~\cite{raffel2020exploring}) using these term-explanation pairs. Third, we extract domain-specific terms from bug reports leveraging two existing methods~\cite{rahman2017strict, tabassum2020code}, explain them using our fine-tuned model, and enrich the bug reports with those explanations. Finally, we conduct extensive experiments using 92,854 bug reports and seven existing baseline methods and examine the impact of our report enrichment (with domain term explanations) on duplicate bug report detection. We not only demonstrate the impact of domain term explanations on seven established techniques but also on textually dissimilar bug reports during duplicate bug report detection. We thus answer two important research questions in this study as follows:

\textbf{(a) RQ$\mathbf{_1}$: Does our enrichment of bug reports with domain term explanations help improve the performance of existing baseline techniques in duplicate bug report detection?}

\looseness=-1
We evaluated our seven baseline techniques spanning five categories using both original and enriched bug reports. Our results show consistent performance improvements across all techniques, with a statistically significant margin in several cases. From traditional Information Retrieval, BM25~\cite{yang2012duplication} showed a 5.16\% improvement in recall for the top 1 result. In the Topic Modeling category, LDA+GloVe~\cite{akilan2020fast} achieved a substantial 41.39\% improvement in recall for the top 5 results. Among Deep Learning based approaches, SiameseCNN~\cite{deshmukh2017towards} achieved a 2.55\% improvement, while DC-CNN~\cite{he2020duplicate} showed a 5.29\% improvement in AUC. BERT-based approaches demonstrated particularly strong improvements, with SBERT~\cite{reimers2019sentence} achieving a 66.88\% improvement for the Recall-rate@1 result and CTEDB~\cite{wu2023intelligent} showing a 5.70\% improvement in Precision. The Large Language Model (LLM) based hybrid approach --CUPID~\cite{zhang2023cupid} achieved a 9.89\% improvement in recall for the top 5 results. These findings provide strong evidence that the enrichment of bug reports enhances the effectiveness of existing duplicate detection significantly across various categories, with particularly notable improvements in BERT-based and Topic Modeling approaches.

\textbf{(b) RQ$\mathbf{_2}$: Does our enrichment of bug reports help the existing baseline techniques in detecting textually dissimilar duplicate bug reports?}

\looseness=-1
We analyze the impact of our bug report enrichment on the detection of textually similar and dissimilar duplicate bug reports. Our experiments reveal that the enrichment leads to improvements in the detection of both types of duplicates by the existing baselines. However, the improvement is much higher with the textually dissimilar duplicate reports. For example, BM25 and LDA+GloVE achieve 55.87\% and 137\% higher recall (i.e., Recall-rate@5) respectively in detecting textually dissimilar duplicates when the bug reports were enriched with the explanations of domain terms. Similarly, the precision improved significantly for DC-CNN, Siamese-CNN, SBERT, CTEDB and CUPID. These results demonstrate that our domain term explanations effectively bridge the semantic gap between duplicate bug reports, especially for textually dissimilar duplicates where the existing techniques often struggle.

\section{Motivating Example} \label{motivation}

\begin{figure}[!t]
  \centering
  \begin{tcolorbox}[boxrule=1pt, colframe=black!60, colback=gray!10, sharp corners, 
  width=\columnwidth, coltitle=white, title= Bug Report \#530801, 
  left=1pt, right=1pt, top=1pt, bottom=1pt] 
  \includegraphics[width=\textwidth, height=\columnwidth, keepaspectratio]{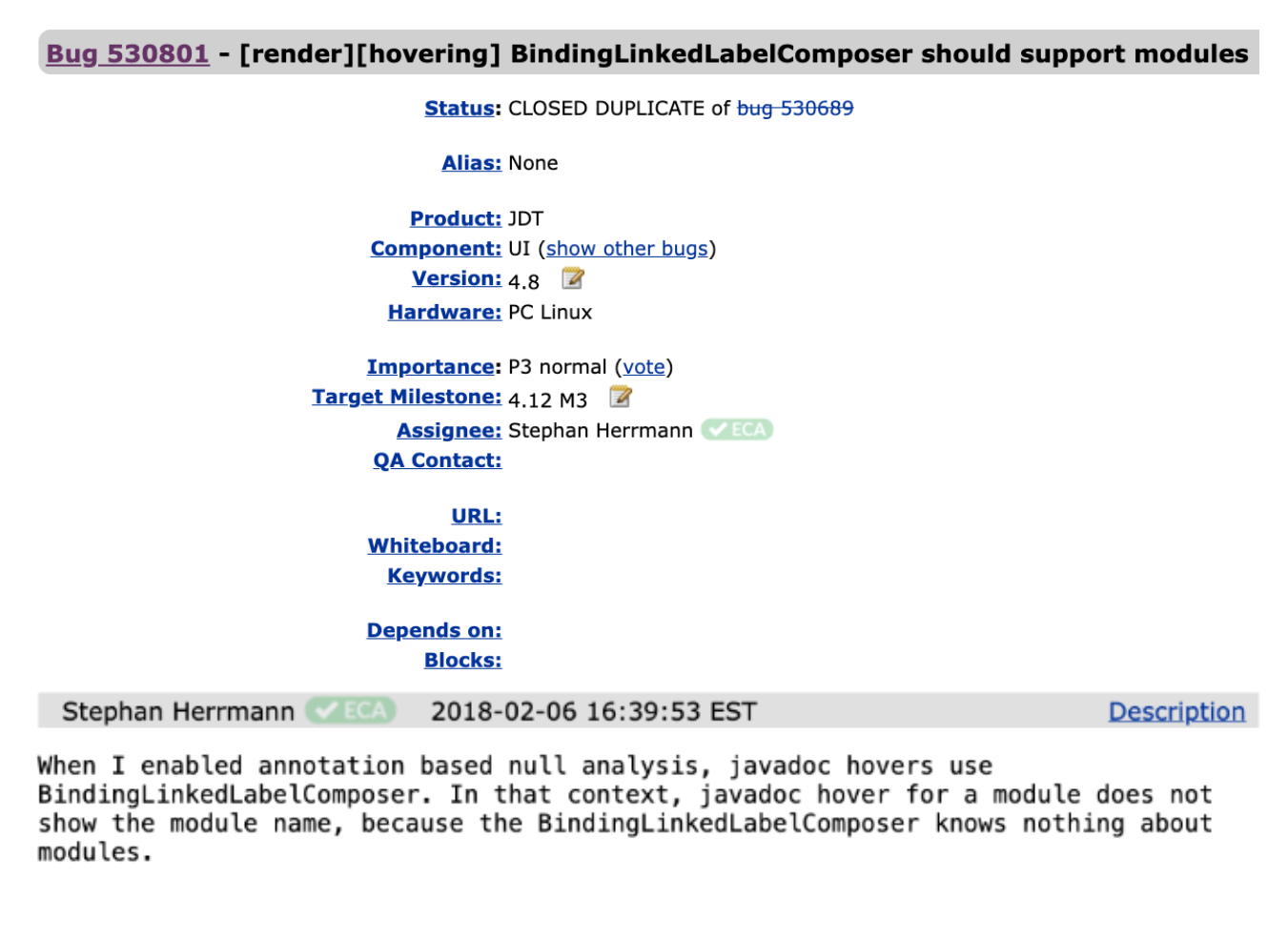} 
  \end{tcolorbox}
  \vspace{-.4cm} 
  \caption{An example bug report from Bugzilla (ID: \#530801) \vspace{-2mm}}
   \vspace{-.4cm}
  \label{fig:motivating_example}
\end{figure}

\looseness=-1
To demonstrate the potential benefits of our work, let us consider the example bug report in Fig.~\ref{fig:motivating_example}. It has been taken from the \emph{Eclipse} project on Bugzilla~\cite{motivating_example}. The example report discusses a bug related to \texttt{BindingLinkedLabelComposer} that lacks awareness of the modules found in an Eclipse project. The problem stems from Javadoc when using the annotation-based null analysis. Specifically, the module name does not appear when the mouse is hovered over a module. The bug report contains a few domain-specific terms or jargon critical to its comprehension. Table~\ref{BugEnricherGeneratedExplanations} shows the domain-specific terms from the bug report as well as their corresponding explanations generated by our fine-tuned T5 model. We leverage these explanations to enrich the example bug report. The enriched bug report can be found in Table~\ref{enriched_bug_report}. 

In our study, we conduct experiments using both these original and the enriched bug reports and found that the rank of their duplicate improved from the 19th to the 13th position in the ranked list when the BM25-based technique was employed~\cite{yang2012duplication,jahan2023towards}. This noticeable improvement in the ranking highlights the benefits of domain term explanations; developers need to check six less bug reports to find the duplicate one. Our empirical study investigates such impacts of domain terms in a comprehensive fashion involving seven baseline techniques and 92.8K bug reports.

\renewcommand{\arraystretch}{1.3}
\begin{table}[htbp]
\centering
\caption{An Enriched Bug Report (ID:\#530801) \vspace{-.3cm} }
\label{enriched_bug_report}
\resizebox{\columnwidth}{!}{%
\begin{tabular}{|l|}
\hline
\rowcolor{black!60} 
\multicolumn{1}{|c|}{\textbf{\textcolor{white}{Enriched Bug Report \#530801}}} \\ \hline \hline
\rowcolor{gray!5}
\begin{tabular}[c]{@{}l@{}} 
When I enabled \textbf{annotation} (\textcolor{black}{\textit{used to describe an annotation object}}) \\ 
based \textbf{null analysis} (\textcolor{black}{\textit{a Java library for analyzing null data}}), \textbf{Javadoc} \\ 
(\textcolor{black}{\textit{documentation generated}}) hovers use \textcolor{black}{\textbf{BindingLinkedLabelComposer}} \\ 
(\textcolor{black}{\textit{used for composing labels}}). In that context, Javadoc hover for a \textbf{module} \\ 
(\textcolor{black}{\textit{a unit of Java code}}) does not show the module name because the \\ 
\textcolor{black}{BindingLinkedLabelComposer} knows nothing about modules. 
\end{tabular} \\ \hline
\end{tabular}
}
\end{table}

\renewcommand{\arraystretch}{1.3}
\begin{table}[htbp]
\centering
\caption{Generated Explanations\vspace{-.3cm}}
\label{BugEnricherGeneratedExplanations}
\resizebox{\columnwidth}{!}{%
\begin{tabular}{ll}
\toprule
\rowcolor{gray!5}
\multicolumn{1}{c}{\textbf{\textcolor{black}{Domain-Specific Term}}} & 
\multicolumn{1}{c}{\textbf{\textcolor{black}{Generated Explanation}}} \\ 
\midrule
\textbf{Javadoc} & \textit{Documentation generated for Java code} \\ 
\rowcolor{gray!5}
\textbf{BindingLinkedLabelComposer} & \textit{Used for composing labels} \\ 
\textbf{Annotation} & \textit{Describes an annotation object} \\ 
\rowcolor{gray!5}
\textbf{Null-analysis} & \textit{Java library for analyzing null data} \\ 
 \textbf{Module} & \textit{A unit of Java code} \\ 
\bottomrule
\end{tabular}}
\end{table}

\section{Study Methodology} \label{methodology}

\begin{figure*}[!t]
  \centering
  \includegraphics[ width=\textwidth, height=8in, keepaspectratio]{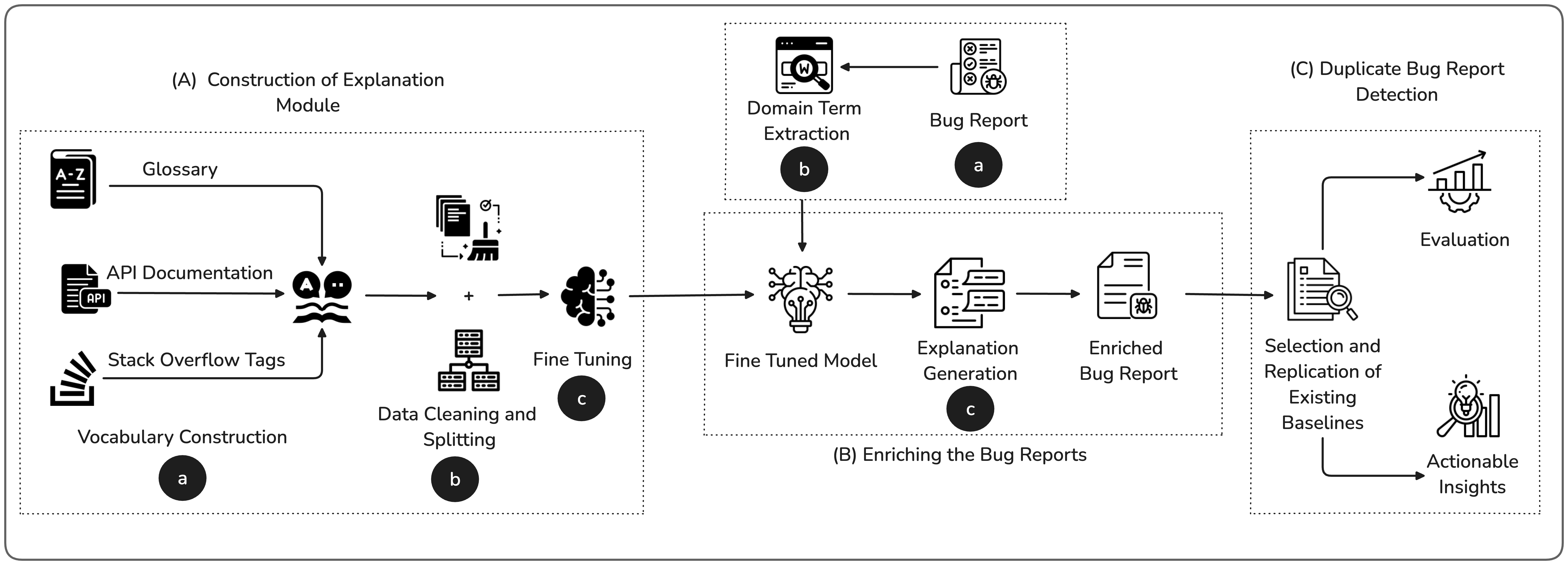}
  \caption{Workflow of our study}
    \vspace{-.3cm}
  \label{fig:schematic}
\end{figure*}

Fig.~\ref{fig:schematic} shows the workflow of our empirical study. First, we construct an explanation module by fine-tuning an LLM (e.g., T5). Second, we enrich bug reports with explanations of their domain terms by employing the fine-tuned model. Finally, we apply the enriched reports to existing baseline techniques to detect duplicate bugs. In this section, we discuss different steps of our study as follows.  

\subsection{Construction of Explanation Module}\label{model}

\textbf{(a) Vocabulary Construction:}
First, we construct a vocabulary of domain-specific terms along with their meanings (a.k.a. explanations) for a popular programming language --- Java. We collect them from three different sources --- StackOverflow Tags, API Documentation and Glossary. 

\textit{StackOverflow Tags:} To collect the domain-specific terms and their meanings, we use StackOverflow as one of our primary sources. Each post on StackOverflow contains several tags that convey the key concepts of the post. The meaning of each tag is defined on StackOverflow, as demonstrated by the example in Table~\ref{sotag2}. We collect 10,022 Java tags from Stack Exchange Data Explorer using the SQL query --- \textsc{select $*$ TagName from Tags}. The ``\texttt{}{Tags}" table contains all the names of the tags in the ``\texttt{TagName}" field. To capture the Java-related tags, we collect a list of Tags that contain the keyword ``\texttt{Java}" in their names. We then scrape the explanations of all the collected Tag names using Beautiful Soup~\cite{bs4}.

\textit{API Documentation:}
\looseness=-1
To gather domain terms, we also collect the API documentation of the most recent stable version of Java (17) (as of August 2024) from their official documentation~\cite{javaDocumentation}. We use Beautiful Soup~\cite{bs4} and Request~\cite{requests}
to scrape the API documentation. 

\renewcommand{\arraystretch}{1.3}
\begin{table}[htbp]
\centering
\caption{A Tag and Explanation at StackOverflow\vspace{-.3cm}}
\label{sotag2}
\resizebox{\columnwidth}{!}{%
\begin{tabular}{|l|}
\hline
\rowcolor{black!60}
\multicolumn{1}{|c|}{\textbf{\textcolor{white}{Tag and Explanation at StackOverflow}}} \\ \hline \hline
\rowcolor{gray!10} 
\textbf{Tag:} \texttt{javafx-11} \\ \hline
\textbf{Explanation:} The JavaFX platform enables developers to create client \\
applications based on JavaSE that behave consistently across
multiple \\ platforms. Built on Java technology since JavaFX 2.0, it was part of \\ the default JDK since JDK 1.8, but starting Java 11, JavaFX is offered \\ as a component separate from the core JDK. \\ \hline
\end{tabular}
}
\end{table}

\looseness=-1
\noindent
First, we scrape the overview page of the official API documentation containing the names of all modules, their explanations, and URLs. We store these module names and their explanations. Then, from the URLs collected in the previous step, we scrape the package and service names, their explanations, and the URLs to corresponding classes and interfaces. Similarly, from the URLs collected in the previous step, we collect the names of the classes and interfaces, their explanations, and the URLs of their fields, methods, and constructors. We also collect the explanations of these class members. In total, we collect 18,738 Java terms and their corresponding explanations from the API documentation. An example term and its explanation from Java 17 are shown in Table~\ref{apitag}.   

\renewcommand{\arraystretch}{1.3}
\begin{table}[htbp]
\centering
\caption{A Term and Explanation from API Documentation\vspace{-.3cm}}
\label{apitag}
\begin{tabular}{|p{0.9\columnwidth}|} 
\hline
\rowcolor{black!60} 
\multicolumn{1}{|c|}{\textbf{\textcolor{white}{Term and Explanation at API Documentation}}} \\ \hline \hline
\rowcolor{gray!10}
\textbf{Term:} \texttt{java.io} \\ \hline
\textbf{Explanation:} Provides for system input and output through \\ data streams,
 serialization and the file system. \\ \hline
\end{tabular}
\end{table}

\renewcommand{\arraystretch}{1.3}
\begin{table}[htbp]
\centering
\caption{A Term and Explanation from Glossary\vspace{-.3cm}}
\label{glossarytag}
\begin{tabular}{|p{0.9\columnwidth}|} 
\hline
\rowcolor{black!60} 
\multicolumn{1}{|c|}{\textbf{\textcolor{white}{Term and Explanation from Java Glossary}}} \\ \hline \hline
\rowcolor{gray!10}
\textbf{Term:} \texttt{immutable} \\ \hline
\textbf{Explanation:} An object with a fixed value. Immutable objects include numbers, strings, and tuples. Such an object cannot be altered. A new object has to be created if a different value has to be stored. They play an important role in places where a constant hash value is needed, for example, as a key in a dictionary. \\ \hline
\end{tabular}
\end{table}
 
\textit{Glossary:}
We also collect 126 Java language-specific terms defined in the glossary, our third source. For Java, we scrape the Oracle's glossary~\cite{java-glossary}.
An example term and explanation from the Java glossary are shown in Table~\ref{glossarytag}. After collecting the data from all three data sources, we discard any duplicates based on exact term matching and keep either the API documentation or Glossary explanations. Since they are both provided by Oracle~\cite{oracle_doc_java17}, the official maintainer of Java, keeping one source of explanation might be sufficient. Table~\ref{tab:dataset} contains the descriptive statistics of our dataset. We find that the average length of each domain-specific term from Java is 13 characters, whereas their explanations have an average length of 188 characters.

\begin{table}[!t]
\caption{Dataset for Domain Terms \& Explanations\vspace{-.3cm}}
\label{tab:dataset}
\resizebox{\columnwidth}{!}{%
\begin{tabular}{|c|c|c|c|c|c|}
    \hline
    \textbf{PL} & \textbf{Source} & \textbf{Size} & \begin{tabular}[c]{@{}c@{}}\textbf{ATL}\\ \textbf{(characters)}\end{tabular} & \begin{tabular}[c]{@{}c@{}}\textbf{AEL}\\ \textbf{(characters)}\end{tabular} & \textbf{Complete Size} \\ \hline \hline
    \multirow{3}{*}{Java} & Stack Overflow & 9,594 & 11.88 & 121.38 & \multirow{3}{*}{21,365} \\ \cline{2-5}
     & API Documentation & 11,771 & 15.90 & 89.61 &  \\ \cline{2-5}
     & Glossary & 244 & 10.87 & 154.19 &  \\ \hline
    \end{tabular}
    }
    \vspace{-0.3cm}
\begin{threeparttable}
\begin{tablenotes}[flushleft]
  \small
  \item \begin{center}
      \item \textbf{PL} $=$ Programming Language, \textbf{ATL} $=$ Average Term Length,
      \item \textbf{AEL} $=$ Average Explanation Length \vspace{-1mm}
  \end{center} 
\end{tablenotes}
\end{threeparttable}
\end{table}

\textbf{ (b) Data Cleaning \& Splitting:}
\looseness=-1
We use standard natural language pre-processing techniques to clean the domain-specific terms or jargon and their explanations. First, we remove the noisy elements --- HTML tags and URLs. Second, we use \textit{``pyspellchecker''}, a spell-checking library~\cite{pyspellchecker}, to correct the spelling of any misspelt words. We then perform lemmatization on all items in our dataset. This step ensures that words are transformed into their root forms, facilitating better analysis~\cite{pramana2022systematic}. After we obtain the cleaned data, we split our dataset (domain terms + explanations) into training, validation and test sets. We split them using the following ratios: 80\% training, 10\% validation, and 10\% testing data~\cite{deshmukh2017towards}. 

\looseness=-1
\textbf{(c) Fine Tuning the T5 Model:}
We then fine-tune an encoder-decoder-based transformer model - T5~\cite{raffel2020exploring} with the collected data from the previous step. The T5 model is trained on the Colossal Clean Crawled Corpus (C4), a large collection of $\approx$750GB of English texts sourced from Common Crawl for text generation tasks. Since pre-training is extremely costly and T5 is already trained on a large amount of English texts, we did not pre-train it with our dataset containing English language texts~\cite{raffel2020exploring}. They can explain partially matched or even lexically unmatched but semantically equivalent domain terms with their strong reasoning capabilities~\cite{tan2023towards}. They also provide context-aware explanations and might be better equipped to tackle the challenges of emerging terminologies due to their large vocabularies, advanced tokenization methods (e.g., Byte-pair encoding~\cite{raffel2020exploring}), and continuous updates by the community (e.g., HuggingFace).

\looseness=-1
\textit{Model Input, Output and Structures:} 
We fine-tune the \texttt{T5-base} variant of the T5 model from HuggingFace~\cite{huggingface_t5} using the \texttt{T5ForCondit\-ionalGeneration} module ~\cite{huggingface_t5}, with our collected data. 
We use domain-specific terms or jargon as input (a.k.a., source sentence) and their explanations as output (a.k.a., target sentence). We train the T5 model with its associated SentencePiece tokenizer~\cite{raffel2020exploring}. The model has a 512-dimensional embedding size, a 6-layer encoder, and eight attention heads per layer. The model also leverages positional embeddings for sequences up to a maximum of 512 tokens. 

\looseness=-1
\textit{Hyperparameter Tuning:} 
Hyperparameter tuning is crucial for optimizing model performance and efficiency, as different datasets and tasks require specific parameter configurations to achieve optimal results~\cite{schratz2019hyperparameter}. Although grid search is a popular method for hyper-parameter tuning, it is not feasible for the T5 model due to its large number of parameters (e.g., 60M to 220M parameters)~\cite{palivela2021optimization}. We thus perform heuristic-based hyperparameter tuning~\cite{mahbub2023explaining}. We adopt an iterative approach and attempt to reach a stable BLEU score on the validation dataset by constantly tuning several parameters from the T5 model, such as learning rate, maximum sequence length, training batch size, and number of training epochs. We also repeat our training with ten random splits of the Java dataset using scikit-learn's library~\cite{scikit} and report the average performance. Based on our hyper-parameter tuning, we chose the following parameters for our model training --- the training and validation batch sizes are both 8; the learning rate is $1e-4$; the maximum source and target text lengths are 128 and 512 tokens, respectively; and a random seed of 42 for reproducibility. Further details about the hyperparameters can be found in the replication package~\cite{bugenricherreplicationpackage}. 

\looseness=-1
\textit{Model Regularization:} Model regularization is critical to prevent overfitting and improve model generalization by controlling complexity and variance in the learned parameters~\cite{goodfellow2016deep}. We adopted the regularization parameters from T5's original architecture~\cite{raffel2020exploring} at HuggingFace, which were found to be useful across various natural language tasks~\cite{phan2021scifive,ranganathan2022text,grover2021deep}. In configuring the model architecture, we set the feed-forward dimension to 2048 and applied a dropout rate of 0.1 for regularization in the model architecture. 

\looseness=-1
\textit{Model Optimization:} Model optimization is essential for efficiently training neural networks by minimizing the loss function. We adjust the model weights to achieve better performance and faster convergence. We use the AdamW optimizer~\cite{loshchilov2017decoupled}, a variant of the Adam optimizer that handles weight decay more effectively.

\subsection{Enriching the Bug Reports}
\looseness=-1
\textbf{(a) Collecting Bug Reports:} To evaluate the impact of domain term explanations on duplicate bug report detection, we first construct our dataset. We collect 92,854 bug reports from an existing benchmark of Jahan et al.~\cite{jahan2023towards} constructed from three open-source systems --- Eclipse, Firefox and Mobile~\cite{jahan2023towards}. We follow the approach of Jahan et al. and apply standard natural language pre-processing to each bug report. We discard stopwords since they do not capture any semantic meaning. We then split the bug report into tokens and removed noisy elements such as non-alphanumeric characters, numbers, HTML tags, and URLs. Lastly, we convert each bug report into lowercase text. 

\begin{figure}[!t]
  \centering
  \includegraphics[ width=\columnwidth, height=10in, keepaspectratio]{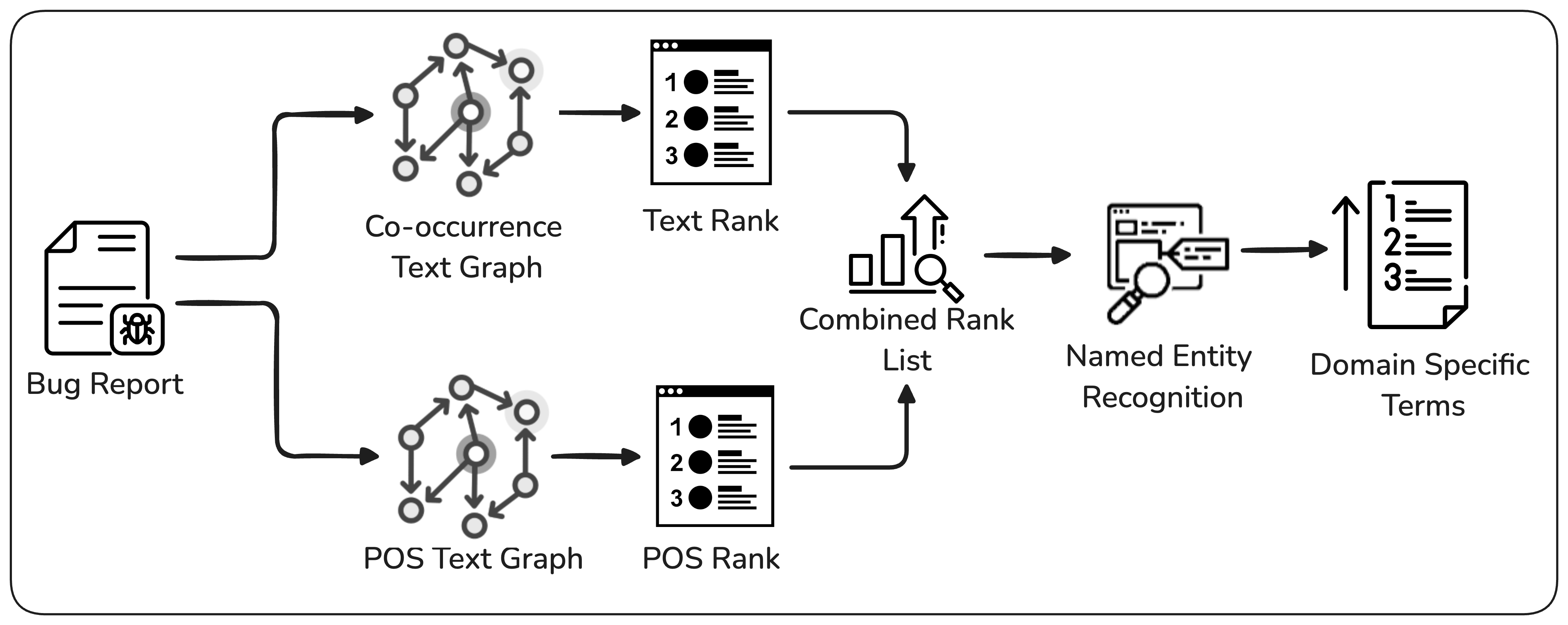}
  \vspace{-.5cm}
  \caption{Domain term extraction }
  \vspace{-.5cm}
\label{fig:domaintermextraction}
\end{figure}

\looseness=-1
\textbf{(b) Domain Term Extraction:}\label{section:domain-term-extraction}
\looseness=-1
Bug reports are written in natural language texts with mentions of various domain-related concepts. Rahman et al.~\cite{rahman2017strict} proposed STRICT that can identify meaningful, domain-related search terms from a bug report leveraging co-occurrences and syntactic relationships among words. On the other hand, BERTOverflow\cite{tabassum2020code} is a custom entity recognition model trained on software engineering corpora. BERTOverflow's training on Stack Overflow posts enables it to accurately identify specialized technical terms like API mentions, programming language constructs, and framework-specific terms that are found in bug reports. Fig.~\ref{fig:domaintermextraction} shows our domain term extraction process. In this work, we leverage term co-occurrences and syntactic relationships from Rahman et al~\cite{rahman2017strict}, and BERTOverflow's NER capabilities to carefully extract the domain-specific terms or jargon from a bug report as follows.

\looseness=-1
We first build a text graph by analyzing the term co-occurrences from each of the sentences in a bug report. In the text graph, each unique term is represented as a vertex, and the co-occurrences of terms in the sentence are represented as the edges among the vertices. We then create another text graph using Parts-of-Speech (POS) dependencies among the terms within the sentences. Then, we use TextRank~\cite{mihalcea2004textrank} to rank the terms in the co-occurrence-based text graph. Similarly, we use POSRank~\cite{blanco2012graph} to rank the terms in the POS-based text graph. Then, we combine the two orthogonal rankings using the Degree of Interest (DOI), as was done by Rahman and Roy~\cite{rahman2016rack}: 
\begin{equation} DOI = \frac{I}{N} \end{equation}
\looseness=-1
Here, $I$ is the position of a domain-specific term in the ranked list, and $N$ is the total number of domain-specific terms. First, we calculate the DOI score of each domain-specific term within the TextRank-based list, and then, we calculate the DOI score within the POSRank-based ranked list. Then, we combine the DOI scores for each domain-specific term and re-rank them based on their combined DOI score using simple summation. Finally, we perform Named Entity Recognition (NER) employing BERTOverflow~\cite{tabassum2020code}, which predicts the relevant technical, domain-specific terms. We then combine these newly identified terms with our previous results and re-rank them using DOI. The above step provides the most important search terms from a bug report (a.k.a., keywords). Once the keywords are available,  we collect the top K (e.g., 10) domain-specific terms/jargon from each bug report with the highest overall score. 

\textbf{(c) Generating Explanations and Enriching Bug Report:}\label{bugenricher:testdataset}
\looseness=-1
We send each of the top K domain-specific terms as input to our fine-tuned T5 model. As input, we provide the following prompt --- ``explain the technical term: [\textit{term}]". The model generates natural language explanations as output, typically 1-2 sentences defining the term and its technical context. For example, given the input ``explain the technical term: \textit{null pointer exception}", the model outputs ``\textit{A programming error that occurs when attempting to use a reference that points to no location in memory (null). This typically causes program crashes during execution}". The model successfully generates explanations for most of the identified terms, with the failures occurring for project-specific abbreviations or very rare technical terms. These explanations are then injected into relevant places within the texts (see Table ~\ref{enriched_bug_report}) to enrich bug reports. We repeat this process for all three subject systems --- Eclipse, Firefox, and Mobile. We call these enriched bug reports --- BR$_{E}$ for subsequent parts of our study. 

\subsection{Duplicate Bug Report Detection} \label{section:replication}
\looseness=-1
\textbf{(a) Replication of existing baselines:} To analyze the impact of bug report enrichment (with domain term explanations) on duplicate bug detection, we replicate seven existing baseline techniques from five frequently used methodologies: Information Retrieval, Topic Modeling, Deep Learning, BERT-based Similarity and Large Language Model. In particular, we select BM25~\cite{yang2012duplication} from Information Retrieval, LDA+GloVe~\cite{akilan2020fast} from Topic Modeling,   SiameseCNN~\cite{deshmukh2017towards}, DC-CNN~\cite{he2020duplicate} from Deep Learning, SBERT~\cite{reimers2019sentence},  CTEDB~\cite{wu2023intelligent} from BERT-based approaches, and CUPID~\cite{zhang2023cupid} from Large Language Model. 

\looseness=-1
BM25 is an established Information Retrieval technique that relies on keyword overlaps between documents. However, it can suffer from the Vocabulary Mismatch Problem (VMP)~\cite{furnas1987vocabulary, chawla2013performance}. LDA+GloVe, a Topic Modeling approach, has the potential to address this limitation. It uses Latent Dirichlet Allocation (LDA) for topic-based clustering, GloVe~\cite{pennington2014glove} for pre-trained word embeddings, and a unified text similarity measure for detecting similar bug reports.
State-of-the-art Deep Learning-based approaches, such as SiameseCNN and DC-CNN, can potentially capture non-linear relationships between variables~\cite{almeida2002predictive, bitvai2015non}. SiameseCNN utilizes a Siamese Convolutional Neural Network architecture, while DC-CNN employs a Deep Convolutional Neural Network for detecting duplicate bug reports. We also replicate two BERT-based techniques -- SBERT~\cite{reimers2019sentence} and CTEDB~\cite{wu2023intelligent}. SBERT leverages Sentence-BERT~\cite{reimers2019sentence} to generate semantically meaningful sentence embeddings that can effectively capture the contextual similarities between bug reports. CTEDB leverages Word2Vec~\cite{church2017word2vec} and TextRank~\cite{mihalcea2004textrank} to extract technical terms from bug reports, utilizes Word2Vec and SBERT for computing semantic similarities, and finally uses DeBERTaV3~\cite{he2021debertav3} to detect the duplicate bug reports. We also replicate -- CUPID~\cite{zhang2023cupid}, a hybrid approach that combines REP~\cite{sun2011towards}, a traditional method for duplicate bug detection, with ChatGPT, a state-of-the-art Large Language Model, to detect the duplicate bug reports. To ensure the reproducibility of our experiments, we utilize the replication packages provided by the authors for LDA+GloVe~\cite{akilan2020fast}, SiameseCNN~\cite{deshmukh2017towards}, DC-CNN~\cite{he2020duplicate}, SBERT~\cite{reimers2019sentence}, and CUPID~\cite{zhang2023cupid}. For BM25, we use the replication package provided by Jahan et al.~\cite{jahan2023towards}. To our knowledge, no replication package is publicly available for CTEDB~\cite{wu2023intelligent}. We thus, replicate it based on the methodology provided by the authors in the paper. Thus, we attempted to mitigate any replication bias using the author-provided replication packages. We also release all relevant artifacts in our replication package~\cite{bugenricherreplicationpackage} for third-party reuse.

\looseness=-1
\textbf{(b) Execution of baseline techniques:}
We conduct experiments using our replicated techniques, on both versions of bug reports - the original bug report (\textit{BR}), and the enriched bug report (\textit{BR$_{E}$}). Then we evaluate their performance in duplicate bug detection using appropriate metrics (Section~\ref{section:evaluationmetrics}). We tested every technique independently on both \textit{BR} and \textit{BR$_{E}$} datasets. To control the confounding factors, we maintain consistent training-testing splits, hyperparameters, and preprocessing steps mentioned in their replication package across all runs, varying only the enrichment factor. Results are captured as rank positions for ranking-based techniques (BM25, LDA, SBERT and CUPID) and binary predictions for classification-based techniques (SiameseCNN, DC-CNN, CTEDB). 

\subsection{Evaluation Metrics} \label{section:evaluationmetrics}
In our experiment, we use five widely used evaluation metrics from existing literature --- Recall Rate@K~\cite{yang2012duplication,akilan2020fast}, Precision~\cite{deshmukh2017towards}, Recall~\cite{deshmukh2017towards}, F1 Score~\cite{deshmukh2017towards} and AUC~\cite{deshmukh2017towards}. To remain consistent with the replicated works, we adopt the same evaluation metrics as the replicated studies. They are defined as follows:  

\subsubsection{\emph{\textit{Recall Rate @ K} }}
\looseness=-1
The Recall-rate@K metric is used to evaluate the effectiveness of duplicate bug report detection techniques. It represents the percentage of bug reports for each of which the corresponding duplicate bug report is found within the top $K$ positions of the ranked results. 
The formula for Recall-rate@K is given by:
\begin{equation}
\text{Recall-rate@K} = \frac{N_{detected}}{N_{total}},
\end{equation}
\looseness=-1
where $N_{detected}$ is the number of bug reports for which the duplicate reports have been correctly identified, and $N_{total}$ is the total number of bug reports being considered. 
We use four different values of $K$ (1, 5, 10, 100) to evaluate the performance of our IR-based and Topic Modeling techniques -- the BM25 and LDA+GloVe models.
\subsubsection{\emph{\textit{Precision} }}
Precision is a metric to assess the correctness of duplicate bug report detection. It represents the percentage of correctly identified duplicate bug report pairs among all detected pairs. 
The formula for precision is given by:
\begin{equation}
\text{Precision} = \frac{TP}{TP + FP},
\end{equation}
where $TP$ stands for true positive (correctly identified duplicate pairs) and $FP$ stands for false positive (incorrectly identified duplicate pairs). 
\subsubsection{\emph{\textit{Recall}}}
Recall measures the percentage of all duplicate bug pairs that are correctly detected by a given technique. 
The formula for recall is given by: 
\begin{equation}
\text{Recall} = \frac{TP}{TP + FN},
\end{equation}
where $TP$ stands for true positive (correctly identified duplicate pairs) and $FN$ stands for false negative (duplicate pairs that were not identified). 

\subsubsection{\emph{\textit{F1 Score}}}
\looseness=-1
The F1 measure is calculated by taking the harmonic mean of precision and recall, as shown in the following formula:
\begin{equation}
\text{F1} = \frac{2 \times \text{Precision} \times \text{Recall}}{\text{Precision} + \text{Recall}}
\end{equation}
The F1 measure combines precision and recall to provide a balanced assessment of the performance in duplicate bug report detection. It helps compare and evaluate different techniques based on their overall performance rather than focusing on just one aspect of their effectiveness. A higher F1 measure indicates a better balance between precision and recall, suggesting that the technique is more effective at identifying duplicate bug reports accurately while minimizing both false positives and false negatives. 

\subsubsection{\emph{\textit{{AUC ---Area Under the Curve}}}}
The Area Under the Curve (AUC) is a performance metric derived from the Receiver Operating Characteristics (ROC) curve, which plots the true positive rate against the false positive rate~\cite{hossin2015review}. AUC represents the fraction of the total area under the ROC curve and ranges from 0 to 1, with a score of 1 indicating perfect classification~\cite{hossin2015review}. Unlike precision and recall, AUC is robust against data imbalance~\cite{davis2006relationship}, making it a more reliable measure of a model's performance when dealing with unequal numbers of positive and negative samples in real-world datasets.

\section{Study Findings} \label{results}

\textit{\textbf{Answering RQ$\mathbf{_1}$} ---  Does our enrichment of bug reports with domain term explanations help improve the performance of existing baseline techniques in duplicate bug report detection?}\label{RQ1}

\looseness=-1
In RQ$\mathbf{_1}$, we analyze how bug report enrichment impacts the performance of seven existing techniques for duplicate bug report detection. Tables~\ref{Table:rq1} and ~\ref{Table:rq1-2} summarize our evaluation details.

\looseness=-1
From Table~\ref{Table:rq1}, we observe that BM25 achieves consistent improvements in Recall-rate@k when using the enriched bug reports (BR$_{E}$). The improvement ranges between 2.5\% to 5\% and is more noticeable for higher values of k. On the other hand, LDA+GloVe achieves more improvements in Recall-rate@k than BM25, especially for higher values of k. For example, at k=5, LDA+GloVe's recall improves by 41.39\% when using the enriched bug reports. While LDA+GloVe shows larger relative improvements, its absolute performance remains lower than its BM25 counterpart, consistent with Jahan et al.'s~\cite{jahan2023towards} findings. The varying impact of enrichment on these techniques stems from their underlying design. BM25's keyword matching benefits from additional technical terms and context, enabling more precise matches. Meanwhile, LDA+GloVe's larger relative improvements, partly due to its lower baseline, demonstrate how enriched reports provide the semantic context needed for the topic models and word embeddings to better capture the relationships between bug reports. Thus, our enrichment strategy successfully improves both lexical matching and semantic analysis during duplicate bug report detection.

\looseness=-1
From Tables~\ref{Table:rq1} and ~\ref{Table:rq1-2}, we observe that state-of-the-art DL techniques -- SiameseCNN, DC-CNN, and BERT-based techniques -- SBERT and CTEDB -- also demonstrate notable improvements when using enriched bug reports. For example, Siamese CNN shows an AUC improvement of 2.55\%, while DC-CNN exhibits a 5.29\% increase in AUC. We observe improvements in Recall, Precision, and F1 scores of up to 5\% for both techniques. Similarly, BERT-based techniques -- SBERT and  CTEDB -- also have gained a boost in performance. For example, SBERT achieves 66.88\% improvement for the Recall-rate@1 and CTEDB achieves 5.70\% improvement in terms of Precision. These consistent improvements across all evaluation metrics suggest that enhancement of bug reports with domain term explanation leads to better representations for bug reports, which in turn enables DL-based and BERT-based techniques to detect duplicate bug reports more accurately. More interestingly, these improvements come with negligible computational overhead. Our enrichment process is computationally efficient, taking less than 3 hours to process the complete dataset of 92.8K reports (i.e., 0.116 seconds per report). This one-time data-preprocessing overhead comprises model inference and term processing and is executed offline. Thus, the computational cost might be negligible, considering the benefits of improved duplicate bug detection.

\renewcommand{\arraystretch}{1.6}
\begin{table}[!t]
\centering
\caption{Performance of Ranking-based Techniques \vspace{-.3cm}}
\label{Table:rq1}
\resizebox{\columnwidth}{!}{%
\begin{tabular}{|ll|cc|cc|cc|cc|}
\hline
\multicolumn{2}{|c|}{\textbf{Technique}} & \multicolumn{2}{c|}{\textbf{BM25}} & \multicolumn{2}{c|}{\textbf{LDA}} & \multicolumn{2}{c|}{\textbf{SBERT}} & \multicolumn{2}{c|}{\textbf{CUPID}} \\ \hline
\multicolumn{2}{|c|}{\textbf{Dataset}} & \multicolumn{1}{c|}{\textbf{\textit{BR}}} & \textbf{\textit{BR$_{E}$}} & \multicolumn{1}{c|}{\textbf{\textit{BR}}} & \textbf{\textit{BR$_{E}$}} & \multicolumn{1}{c|}{\textbf{\textit{BR}}} & \textbf{\textit{BR$_{E}$}} & \multicolumn{1}{c|}{\textbf{\textit{BR}}} & \textbf{\textit{BR$_{E}$}} \\ \hline \hline
\multicolumn{1}{|l|}{\multirow{4}{*}{Recall-rate@k}} & k=1 & \multicolumn{1}{c|}{18.67} & \textbf{19.63} & \multicolumn{1}{c|}{0.00} & \textbf{0.00} & \multicolumn{1}{c|}{12.98} & \textbf{21.65} & \multicolumn{1}{c|}{19.63} & \textbf{20.50} \\ \cline{2-10} 
\multicolumn{1}{|l|}{} & k=5 & \multicolumn{1}{c|}{31.18} & \textbf{31.85} & \multicolumn{1}{c|}{5.50} & \textbf{7.78} & \multicolumn{1}{c|}{24.49} & \textbf{28.62} & \multicolumn{1}{c|}{31.85} & \textbf{35.00} \\ \cline{2-10} 
\multicolumn{1}{|l|}{} & k=10 & \multicolumn{1}{c|}{37.02} & \textbf{38.39} & \multicolumn{1}{c|}{9.33} & \textbf{13.17} & \multicolumn{1}{c|}{29.94} & \textbf{35.37} & \multicolumn{1}{c|}{38.39} & \textbf{42.05} \\ \cline{2-10} 
\multicolumn{1}{|l|}{} & k=100 & \multicolumn{1}{c|}{55.86} & \textbf{57.39} & \multicolumn{1}{c|}{17.50} & \textbf{20.01} & \multicolumn{1}{c|}{49.13} & \textbf{57.27} & \multicolumn{1}{c|}{80.50} & \textbf{84.20} \\ \hline
\end{tabular}}
\end{table}

\begin{table}[!t]
\centering
\caption{Performance of Classification-based Techniques\vspace{-.3cm}}
\label{Table:rq1-2}
\resizebox{0.8\columnwidth}{!}{%
\begin{tabular}{|l|cc|cc|ll|}
\hline
\multicolumn{1}{|c|}{\textbf{Technique}} & \multicolumn{2}{c|}{\textbf{SiameseCNN}} & \multicolumn{2}{c|}{\textbf{DC-CNN}} & \multicolumn{2}{c|}{\textbf{CTEDB}} \\ \hline
\textbf{Dataset} & \multicolumn{1}{c|}{\textbf{\textit{BR}}} & \textbf{\textit{BR$_{E}$}} & \multicolumn{1}{c|}{\textbf{\textit{BR}}} & \textbf{\textit{BR$_{E}$}} & \multicolumn{1}{c|}{\textbf{\textit{BR}}} & \multicolumn{1}{c|}{\textbf{\textit{BR$_{E}$}}} \\ \hline \hline
AUC & \multicolumn{1}{c|}{70.30} & \textbf{72.09} & \multicolumn{1}{c|}{89.18} & \textbf{93.90} & \multicolumn{1}{l|}{87.12} & \textbf{91.17} \\ \hline
Recall & \multicolumn{1}{c|}{89.67} & \textbf{90.37} & \multicolumn{1}{c|}{89.15} & \textbf{92.85} & \multicolumn{1}{l|}{87.42} & \textbf{90.74 } \\ \hline
Precision & \multicolumn{1}{c|}{87.67} & \textbf{88.45} & \multicolumn{1}{c|}{88.14} & \textbf{90.56} & \multicolumn{1}{l|}{90.49} &  \textbf{95.65}\\ \hline
F1 & \multicolumn{1}{c|}{88.64} & \textbf{89.39} & \multicolumn{1}{c|}{88.83} & \textbf{91.13} & \multicolumn{1}{l|}{89.43} & \textbf{91.60} \\ \hline  
\end{tabular}  }
\begin{threeparttable}
\begin{tablenotes}[flushleft]
  \small
  \item \begin{center}
      \item  \vspace{-2mm}  \textbf{BR} $=$ Bug Reports, \textbf{BR$_{E}$ } $=$ Bug Reports Enriched,
  \end{center} 
\end{tablenotes}
\end{threeparttable}
\vspace{-0.2cm}
\end{table}

\renewcommand{\arraystretch}{1.3}
\begin{table}[!t]
\caption{Statistical Significance of Existing Techniques\vspace{-.3cm} }
\label{Table:rq1stats}
\centering
\resizebox{0.8\columnwidth}{!}{%
\begin{tabular}{|l|ccc|}
\hline
\multicolumn{1}{|c|}{\multirow{2}{*}{\textbf{Method}}} & \multicolumn{3}{c|}{\textbf{Complete Dataset}} \\ \cline{2-4} 
\multicolumn{1}{|c|}{} & \multicolumn{1}{c|}{\textbf{\begin{tabular}[c]{@{}c@{}}Probability \\ Distribution\end{tabular}}} & \multicolumn{1}{c|}{\textbf{\begin{tabular}[c]{@{}c@{}}Signifiance \\ (p-value)\end{tabular}}} & \textbf{Effect Size} \\ \hline \hline
BM25 & \multicolumn{1}{c|}{N} & \multicolumn{1}{c|}{1.86E-15} & (Small) 0.1853 \\ \hline
LDA+GloVe & \multicolumn{1}{c|}{NN} & \multicolumn{1}{c|}{8.86E-05} & (Large) 0.5896 \\ \hline \hline
SiameseCNN & \multicolumn{1}{c|}{N} & \multicolumn{1}{c|}{8.12E-04} & (Small) 0.2132 \\ \hline
DC-CNN & \multicolumn{1}{c|}{N} & \multicolumn{1}{c|}{1.80E-10} & (Large) 3.9131 \\ \hline
SBERT & \multicolumn{1}{c|}{NN} & \multicolumn{1}{c|}{9.54E-07} & (Large) 0.5465 \\ \hline
CTEDB & \multicolumn{1}{c|}{N} & \multicolumn{1}{c|}{1.15E-06} & (Large) 2.4194 \\ \hline \hline
CUPID & \multicolumn{1}{c|}{NN} & \multicolumn{1}{c|}{0.0098} & (Large) 0.0159 \\ \hline 
\end{tabular}}\vspace{-2mm} 
\begin{threeparttable}
\begin{tablenotes}[flushleft]
  \small
  \item \begin{center}
      \item \textbf{N} $=$ Normal Distribution, \textbf{NN} $=$ Non Normal Distribution
  \end{center} \vspace{-3mm} 
\end{tablenotes}
\end{threeparttable}
\end{table}

\looseness=-1
From Tables~\ref{Table:rq1} and ~\ref{Table:rq1-2}, we observe that the LLM-based hybrid approach -- CUPID -- also demonstrates notable improvements when using enriched bug reports. CUPID achieves a noticeable performance improvement in Recall-rate@k ranging between 4.5\% to 9.89\%. This improvement suggests that enriched bug reports (with domain term explanations) provide better quality input for both the traditional REP component and the ChatGPT component of CUPID, leading to more accurate duplicate detection. Like other techniques, these improvements are also achieved with minimal computational overhead during the enrichment process. We also conducted statistical significance tests to determine whether the performance improvement of the baseline techniques was significant or not. For ranking-based techniques (BM25, LDA, SBERT and CUPID), we evaluated their performance by varying k values from 1 to 100 (with an interval of 5) using both original and enriched bug reports~\cite{jahan2023towards}. For classification-based techniques (SiameseCNN, DC-CNN, CTEDB), we collect 20 random sub-samples of the test dataset and measured the technique's performance on each subsample~\cite{arcuri2011practical}.

\looseness=-1
We then conduct \textit{Shapiro-Wilk normality test}~\cite{royston1992approximating} on the results from all seven techniques. For each technique, we collected the results using both the original bug reports and their enriched versions. For normal distribution, we perform \textit{paired t-test}~\cite{kim2015t}, and for non-normal distribution, we perform \textit{Wilcoxon signed‐rank test}~\cite{woolson2005wilcoxon}. Furthermore, we also calculate the effect size to determine the extent of the significance. For normal distribution, we use \textit{Cohen's D test}~\cite{gignac2016effect}, and for non-normal distribution, we use \textit{Cliff's Delta test}~\cite{macbeth2011cliff}. Table~\ref{Table:rq1stats} summarizes our analysis from the statistical tests. We find that the performance improves by a statistically significant margin for all techniques, with a small effect on BM25 and SiameseCNN and a large effect on LDA+GloVe, DC-CNN, SBERT, CTEDB and CUPID. These effect sizes indicate that our enrichment approach significantly impacts all techniques, with DC-CNN showing the largest effect size. In other words, while all techniques benefit from the enriched bug reports, DC-CNN's sophisticated architecture can better utilize the enhanced semantic information. 

\looseness=-1
The significant performance gains above by different techniques can be attributed to the enriched semantic context provided by our domain term explanations. When technical terms and API names in bug reports are enriched with their definitions, they create a richer representation that helps bridge the vocabulary gap between bug reports. However, different baseline techniques, empowered by different methodologies, were able to leverage such enhanced representations to varying extents, as shown by our experiments.

\renewcommand{\arraystretch}{1.1}
\begin{table}[!t]
\centering
\caption{Evaluation of Explanation\vspace{-.3cm}}
\label{Table:explanation-quality}
\small 
\resizebox{0.8\columnwidth}{!}{
\begin{tabular}{|c|c|c|c|}
\hline
\textbf{Model} & \textbf{BLEU} & \textbf{METEOR} & \textbf{Semantic Similarity} \\ \hline \hline
\begin{tabular}[c]{@{}c@{}}T5$_{Java}$\end{tabular} & 28.85 & 0.27 & 53.26 \\ \hline
\end{tabular}} 
\end{table}

\begin{figure}[]
  \centering
  \includegraphics[width=\columnwidth, keepaspectratio]{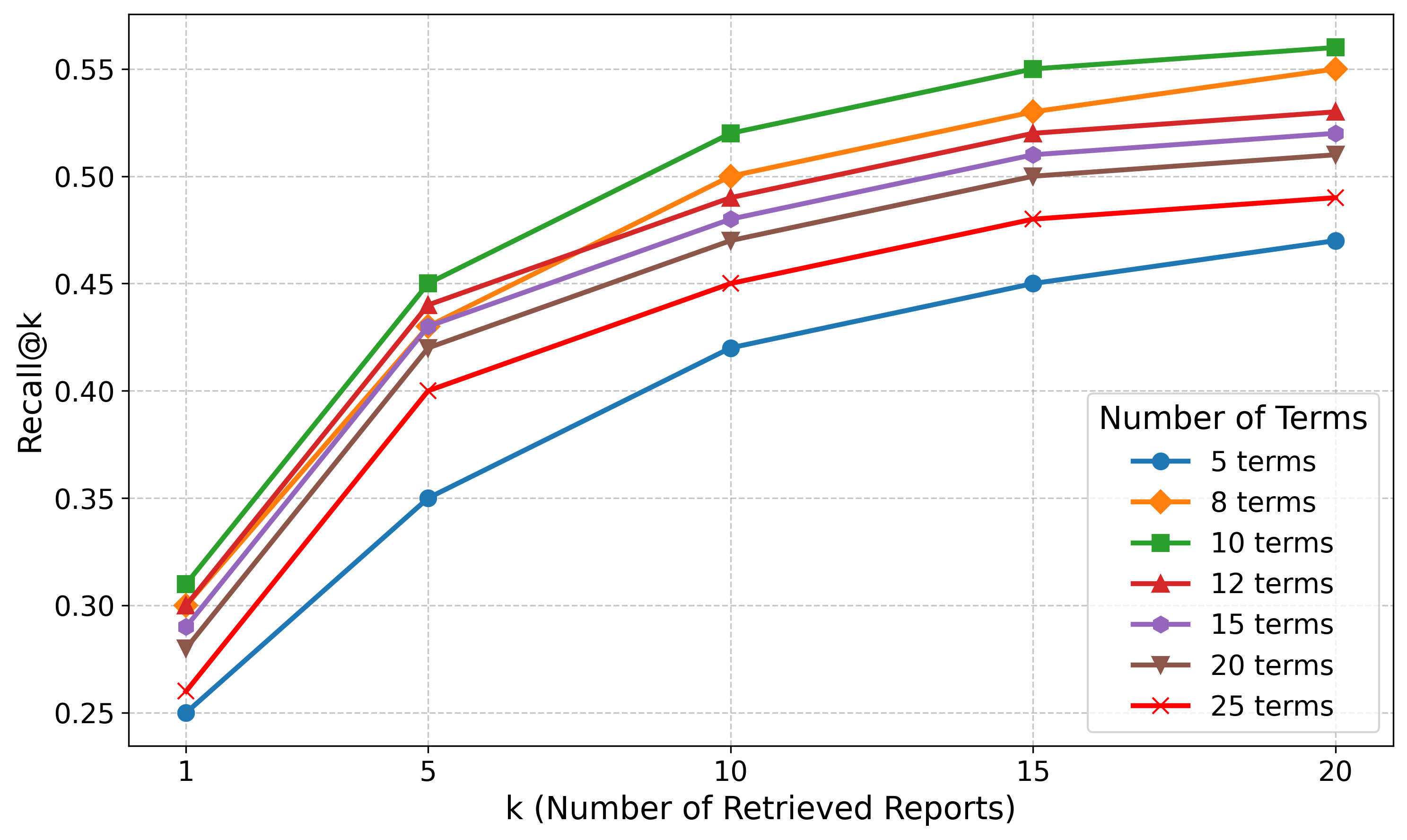}
  \caption{Impact of the number of domain term explanations \vspace{-0.9cm}}
  \label{fig:domain-term-impact}
\end{figure}

\looseness=-1
We also analyze how the number of explained domain terms impacts the performance of existing baseline techniques. Here, we select BM25 due to its simple methodology and significant performance gain. By adopting our bug report enrichment (Section ~\ref{section:domain-term-extraction}), we extract and explain 5 to 25 domain terms from each bug report and evaluate BM25 in duplicate bug detection. Fig. 4 summarizes our evaluation details. We find that 10 terms provide the optimal balance between performance convergence and effectiveness of report enrichment. As shown in Figure~\ref{fig:domain-term-impact}, Recall@K improves with increased terms but reaches the maximum with 10 terms. In other words, the terms ranked beyond the top 10 position possibly introduce noise whereas less than 10 terms might fail to provide enough information to enrich a bug report during duplicate detection. 

\looseness=-1
We also assess the quality of our generated explanations for domain terms from bug reports. We evaluate them using ground truth explanations (Section ~\ref{model}) and three evaluation metrics --BLEU~\cite{papineni2002bleu}, METEOR~\cite{banerjee2005meteor} and Semantic Similarity (SS)~\cite{haque2022semantic}. As shown in Table ~\ref{Table:explanation-quality}, our explanations achieve a BLEU score of 28.85, suggesting that they are understandable or good, according to Google’s Translation standards~\cite{automldoc}. While BLEU focuses on precision, we also use the METEOR score, which incorporates synonyms, word forms, and sentence structure to better capture recall~\cite{banerjee2005meteor}. The METEOR score of 0.27 indicates the model's ability to produce significant parts of the ground truth. However, since BLEU and METEOR rely on keyword matching, they may not fully capture semantic relevance. Thus, we evaluate using Semantic Similarity, achieving a maximum of 53.26\%, indicating substantial semantic overlap with the ground truth. These results demonstrate that our generated domain-term explanations achieve strong syntactic and semantic alignment with the ground truth explanations.

\begin{boxH}
\looseness=-1
\textbf{Summary of RQ1:} 
Enriched bug reports lead to improved performance in duplicate bug report detection across all detection techniques, with SBERT achieving the highest gain in recall (66.88\% for Recall-rate@1), LDA+GloVe showing the best recall improvement (41.39\% for Recall-rate@5), and DC-CNN increasing AUC by 5.29\%. CUPID improved Recall-rate@5 performance by up to 9.89\%. Statistical tests suggest significance in performance improvements, revealing large effect sizes for most techniques, with DC-CNN exhibiting the largest effect size (3.9131).
\end{boxH}

\renewcommand{\arraystretch}{1.6}
\begin{table}[]
\caption{Performance of Ranking-based Techniques\vspace{-.3cm}}
\label{Table:rq2}
\centering
\resizebox{\columnwidth}{!}{%
\begin{tabular}{|c|l|cccccccc|}
\hline
\multirow{3}{*}{\textbf{Technique}} & \multicolumn{1}{c|}{\multirow{3}{*}{\textbf{Dataset}}} & \multicolumn{8}{c|}{\textbf{Recall-rate@k}} \\ \cline{3-10} 
 & \multicolumn{1}{c|}{} & \multicolumn{4}{c|}{\textbf{Textually Similar}} & \multicolumn{4}{c|}{\textbf{Textually Dissimilar}} \\ \cline{3-10} 
 & \multicolumn{1}{c|}{} & \multicolumn{1}{c|}{\textbf{k=1}} & \multicolumn{1}{c|}{\textbf{k=5}} & \multicolumn{1}{c|}{\textbf{k=10}} & \multicolumn{1}{c|}{\textbf{k=100}} & \multicolumn{1}{c|}{\textbf{k=1}} & \multicolumn{1}{c|}{\textbf{k=5}} & \multicolumn{1}{c|}{\textbf{k=10}} & \textbf{k=100} \\ \hline \hline
\multirow{2}{*}{BM25} & \textit{BR} & \multicolumn{1}{c|}{22.29} & \multicolumn{1}{c|}{38.71} & \multicolumn{1}{c|}{43.73} & \multicolumn{1}{c|}{62.72} & \multicolumn{1}{c|}{16.40} & \multicolumn{1}{c|}{28.80} & \multicolumn{1}{c|}{34.60} & 52.95 \\ \cline{2-10} 
 & \textit{\textbf{BR$_{E}$}} & \multicolumn{1}{c|}{\textbf{28.45}} & \multicolumn{1}{c|}{\textbf{42.84}} & \multicolumn{1}{c|}{\textbf{47.96}} & \multicolumn{1}{c|}{\textbf{65.95}} & \multicolumn{1}{c|}{\textbf{25.56}} & \multicolumn{1}{c|}{\textbf{31.99}} & \multicolumn{1}{c|}{\textbf{37.16}} & \textbf{57.36} \\ \hline
\multirow{2}{*}{LDA} & \textit{BR} & \multicolumn{1}{c|}{0.00} & \multicolumn{1}{c|}{6.67} & \multicolumn{1}{c|}{9.83} & \multicolumn{1}{c|}{18.50} & \multicolumn{1}{c|}{0.00} & \multicolumn{1}{c|}{3.67} & \multicolumn{1}{c|}{7.00} & 13.83 \\ \cline{2-10} 
 & \textit{\textbf{BR$_{E}$}} & \multicolumn{1}{c|}{\textbf{4.14}} & \multicolumn{1}{c|}{\textbf{10.88}} & \multicolumn{1}{c|}{\textbf{15.39}} & \multicolumn{1}{c|}{\textbf{22.42}} & \multicolumn{1}{c|}{\textbf{3.45}} & \multicolumn{1}{c|}{\textbf{8.69}} & \multicolumn{1}{c|}{\textbf{13.37}} & \textbf{19.30} \\ \hline
\multirow{2}{*}{S-BERT} & \textit{BR} & \multicolumn{1}{c|}{53.97} & \multicolumn{1}{c|}{76.09} & \multicolumn{1}{c|}{79.88} & \multicolumn{1}{c|}{92.29} & \multicolumn{1}{c|}{52.91} & \multicolumn{1}{c|}{74.60} & \multicolumn{1}{c|}{78.31} & 90.48 \\ \cline{2-10} 
 & \textit{\textbf{BR$_{E}$}} & \multicolumn{1}{c|}{\textbf{56.84}} & \multicolumn{1}{c|}{\textbf{86.16}} & \multicolumn{1}{c|}{\textbf{88.36}} & \multicolumn{1}{c|}{\textbf{96.04}} & \multicolumn{1}{c|}{\textbf{54.13}} & \multicolumn{1}{c|}{\textbf{82.06}} & \multicolumn{1}{c|}{\textbf{82.64}} & \textbf{93.38} \\ \hline
\multirow{2}{*}{CUPID} & \textit{BR} & \multicolumn{1}{c|}{28.45} & \multicolumn{1}{c|}{42.84} & \multicolumn{1}{c|}{47.96} & \multicolumn{1}{c|}{78.50} & \multicolumn{1}{c|}{25.56} & \multicolumn{1}{c|}{31.99} & \multicolumn{1}{c|}{37.16} & 71.50 \\ \cline{2-10} 
 & \textit{\textbf{BR$_{E}$}} & \multicolumn{1}{c|}{\textbf{29.23}} & \multicolumn{1}{c|}{\textbf{46.34}} & \multicolumn{1}{c|}{\textbf{48.97}} & \multicolumn{1}{c|}{\textbf{85.49}} & \multicolumn{1}{c|}{\textbf{27.00}} & \multicolumn{1}{c|}{\textbf{34.50}} & \multicolumn{1}{c|}{\textbf{40.50}} & \textbf{75.33} \\ \hline
\end{tabular}
}
\begin{threeparttable}
\end{threeparttable}
\end{table}
\renewcommand{\arraystretch}{1.6}

\begin{table}[]
\caption{Performance of Classification-based Techniques\vspace{-.3cm}}
\label{Table:rq2-2}
\centering
\resizebox{\columnwidth}{!}{%
\begin{tabular}{|c|l|cccc|cccc|}
\hline
\multirow{2}{*}{\textbf{Technique}} & \multicolumn{1}{c|}{\multirow{2}{*}{\textbf{Dataset}}} & \multicolumn{4}{c|}{\textbf{Textually Similar}} & \multicolumn{4}{c|}{\textbf{Textually Dissimilar}} \\ \cline{3-10} 
 & \multicolumn{1}{c|}{} & \multicolumn{1}{c|}{\textbf{AUC}} & \multicolumn{1}{c|}{\textbf{Recall}} & \multicolumn{1}{c|}{\textbf{Precision}} & \textbf{F1} & \multicolumn{1}{c|}{\textbf{AUC}} & \multicolumn{1}{c|}{\textbf{Recall}} & \multicolumn{1}{c|}{\textbf{Precision}} & \textbf{F1} \\ \hline \hline
\multirow{2}{*}{SiameseCNN} & \textit{BR} & \multicolumn{1}{c|}{62.24} & \multicolumn{1}{c|}{59.67} & \multicolumn{1}{c|}{72.67} & 65.31 & \multicolumn{1}{c|}{54.93} & \multicolumn{1}{c|}{50.33} & \multicolumn{1}{c|}{62.00} & 55.51 \\ \cline{2-10} 
 & \textit{\textbf{BR$_{E}$}} & \multicolumn{1}{c|}{\textbf{63.08}} & \multicolumn{1}{c|}{\textbf{60.98}} & \multicolumn{1}{c|}{\textbf{73.71}} & \textbf{66.54} & \multicolumn{1}{c|}{\textbf{56.14}} & \multicolumn{1}{c|}{\textbf{51.57}} & \multicolumn{1}{c|}{\textbf{63.17}} & \textbf{56.73} \\ \hline
\multirow{2}{*}{DC-CNN} & \textit{BR} & \multicolumn{1}{c|}{65.12} & \multicolumn{1}{c|}{56.89} & \multicolumn{1}{c|}{73.56} & 68.13 & \multicolumn{1}{c|}{54.67} & \multicolumn{1}{c|}{51.45} & \multicolumn{1}{c|}{59.79} & 56.28 \\ \cline{2-10} 
 & \textit{\textbf{BR$_{E}$}} & \multicolumn{1}{c|}{\textbf{69.91}} & \multicolumn{1}{c|}{\textbf{61.45}} & \multicolumn{1}{c|}{\textbf{76.89}} & \textbf{70.72} & \multicolumn{1}{c|}{\textbf{57.39}} & \multicolumn{1}{c|}{\textbf{55.68}} & \multicolumn{1}{c|}{\textbf{65.24}} & \textbf{60.34} \\ \hline
\multirow{2}{*}{CTEDB} & \textit{BR} & \multicolumn{1}{c|}{70.14} & \multicolumn{1}{c|}{63.5} & \multicolumn{1}{c|}{69.16} & 66.2 & \multicolumn{1}{c|}{59.47} & \multicolumn{1}{c|}{53.12} & \multicolumn{1}{c|}{66.29} & 58.97 \\ \cline{2-10} 
 & \textit{\textbf{BR$_{E}$}} & \multicolumn{1}{c|}{\textbf{72.8}} & \multicolumn{1}{c|}{\textbf{65.33}} & \multicolumn{1}{c|}{\textbf{71.39}} & \textbf{68.22} & \multicolumn{1}{c|}{\textbf{63.69}} & \multicolumn{1}{c|}{\textbf{57.72}} & \multicolumn{1}{c|}{\textbf{71.36}} & \textbf{63.81} \\ \hline
\end{tabular}
}
\begin{threeparttable}
\begin{tablenotes}[flushleft]
  \small
  \item \begin{center}
      \item \vspace{-4mm}\textbf{BR} $=$ Bug Reports, \textbf{BR$_{E}$ } $=$ Bug Reports Enriched \vspace{-4mm}
  \end{center} 
\end{tablenotes}
\end{threeparttable}
\end{table}

\noindent
\textit{\textbf{Answering RQ$\mathbf{_2}$} --- Does our enrichment of bug reports help the existing baseline techniques in detecting textually dissimilar duplicate bug reports?}\label{RQ2}

\looseness=-1
In RQ2, we investigate whether our enrichment of bug reports can improve the detection of textually dissimilar duplicate bug reports, a challenging problem during bug resolution. Towards this end, we analyze the performance of seven existing techniques against textually similar and textually dissimilar bug reports, collected from the benchmark dataset of Jahan et.~\cite {jahan2023towards}. Tables~\ref{Table:rq2} and ~\ref{Table:rq2-2} summarizes our investigation details.

From Table~\ref{Table:rq2}, we see that BM25 and LDA+GloVe achieve consistent improvements in Recall-rate@k due to report enhancement for both textually similar and dissimilar duplicate reports. In the case of BM25, the improvements are more pronounced for textually dissimilar duplicates. For example, at k=1, the recall improves by 55.87\% for textually dissimilar duplicates, while it improves by 27.62\% for textually similar duplicates. Similarly, LDA+GloVe shows substantial improvements in Recall-rate@k for both textually similar and dissimilar duplicates when using the enriched dataset. For example, at k=5, the recall improves by 137\% for textually dissimilar duplicates, while it improves by 63.15\% for textually similar duplicates. All these statistics above suggest that our bug report enhancement using domain term explanation was found to be effective when detecting textually dissimilar duplicates using the existing techniques -- BM25 and LDA+GloVe.

\looseness=-1
Tables~\ref{Table:rq2} and ~\ref{Table:rq2-2}, we see that DL techniques -- SiameseCNN, DC-CNN, and BERT-based techniques -- SBERT and CTEDB achieve performance gains when detecting textually similar and dissimilar duplicate bug reports. For example, SiameseCNN achieves performance gains of 1.88\% in terms of AUC, 2.46\% in terms of Recall and 2.19\% in terms of F1 score when detecting textually dissimilar duplicates, enriched by our domain term explanations. It also achieves Precision improvement by 9.12\% higher precision for textually dissimilar duplicates compared to 4.53\% for textually similar duplicates. BERT-based techniques -- SBERT and CTEDB -- also show similar trends. For example, CTEDB achieves an improvement in recall of 8.66\% for textually dissimilar duplicates compared to 2.88\% for textually similar duplicates. SBERT, on the other hand, achieves a performance improvement of 13.24\% at k=5 for textually similar duplicates compared to 10\% for textually dissimilar duplicates. Thus, performance improvements for SiameseCNN, DC-CNN and CTEDB were notably higher for textually dissimilar duplicates, ranging from 2.5\%-9\% compared to textually similar ones ranging from 1\%-7\%. On the other hand, SBERT shows a comparable performance improvement for both textually similar and dissimilar duplicates. All these results indicate that our enrichment of bug reports using domain term explanations positively influences the deep learning techniques in detecting both textually similar and dissimilar duplicates. 
\renewcommand{\arraystretch}{2}
\begin{table}[!t]
\caption{Statistical Significance on BR and BR$_{E}$\vspace{-.3cm}}
\centering
\resizebox{\columnwidth}{!}{%
\label{Table:rq2stats}
\begin{tabular}{|l|ccc|ccc|}
\hline
\multicolumn{1}{|c|}{\multirow{2}{*}{\textbf{Method}}} & \multicolumn{3}{c|}{\textbf{Textually Similar}} & \multicolumn{3}{c|}{\textbf{Textually Dissimilar}} \\ \cline{2-7} 
\multicolumn{1}{|c|}{} & \multicolumn{1}{c|}{\textbf{\begin{tabular}[c]{@{}c@{}}Probability \\ Distribution\end{tabular}}} & \multicolumn{1}{c|}{\textbf{\begin{tabular}[c]{@{}c@{}}Signifiance \\ (p-value)\end{tabular}}} & \textbf{Effect Size} & \multicolumn{1}{c|}{\textbf{\begin{tabular}[c]{@{}c@{}}Probability \\ Distribution\end{tabular}}} & \multicolumn{1}{c|}{\textbf{\begin{tabular}[c]{@{}c@{}}Signifiance \\ (p-value)\end{tabular}}} & \textbf{Effect Size} \\ \hline \hline
BM25 & \multicolumn{1}{c|}{NN} & \multicolumn{1}{c|}{9.54E-07} & (Small) 0.2834 & \multicolumn{1}{c|}{N} & \multicolumn{1}{c|}{6.46E-07} & (Small) 0.3106 \\ \hline
LDA+GloVe & \multicolumn{1}{c|}{N} & \multicolumn{1}{c|}{5.57E-21} & (Medium) 0.6644 & \multicolumn{1}{c|}{N} & \multicolumn{1}{c|}{9.01E-15} & (Large) 1.1859 \\ \hline \hline
SiameseCNN & \multicolumn{1}{c|}{NN} & \multicolumn{1}{c|}{0.03999} & (Small) 0.4175 & \multicolumn{1}{c|}{NN} & \multicolumn{1}{c|}{0.00444} & (Large) 1.0866 \\ \hline
DC-CNN & \multicolumn{1}{c|}{NN} & \multicolumn{1}{c|}{1.91E-06} & (Large) 1.0000 & \multicolumn{1}{c|}{N} & \multicolumn{1}{c|}{2.09E-08} & (Large) 2.4096 \\ \hline
SBERT & \multicolumn{1}{c|}{NN} & \multicolumn{1}{c|}{0.02157} & (Small) 0.3197 & \multicolumn{1}{c|}{NN} & \multicolumn{1}{c|}{6.18E-05} & (Medium) 0.3197 \\ \hline
CTEDB & \multicolumn{1}{c|}{N} & \multicolumn{1}{c|}{0.00056} & (Large) 1.1151 & \multicolumn{1}{c|}{N} & \multicolumn{1}{c|}{4.51E-07} & (Large) 2.4331 \\ \hline \hline
CUPID & \multicolumn{1}{c|}{NN} & \multicolumn{1}{c|}{0.0077} & (Small) 0.0249 & \multicolumn{1}{c|}{NN} & \multicolumn{1}{c|}{9.54E-07} & (Small) 0.2132 \\ \hline
\end{tabular}
}\vspace{-1mm}
\begin{threeparttable}
\begin{tablenotes}[flushleft]
  \small
  \item \begin{center}
      \item \textbf{N} $=$ Normal Distribution, \textbf{NN} $=$ Non Normal Distribution \vspace{-.5cm}
  \end{center} 
\end{tablenotes}
\end{threeparttable}
\end{table}

Tables~\ref{Table:rq2} show the performance of CUPID on textually similar and dissimilar duplicate bug reports. CUPID demonstrates performance improvements across both datasets when using enriched bug reports. For example, at k = 10, recall improves by 9\% for textually dissimilar duplicates compared to 2.11\% for textually similar duplicates. CUPID shows an overall performance gain ranging from  5\%-9\% for textually dissimilar compared to 2\%-8\% for textually dissimilar duplicates. This pattern aligns with our observations from other techniques, where the enrichment process particularly benefits the detection of textually dissimilar duplicates. 

\looseness=-1
We also conduct statistical significance tests to determine whether the performance improvement is significant across the seven existing techniques for textually similar and dissimilar duplicate bug reports. We follow the same statistical test methodology as RQ$\mathbf{_1}$ for consistency. Table~\ref{Table:rq2stats} provides the complete test statistics and analysis. For both datasets, we find statistically significant performance improvements, with higher effect sizes for textually dissimilar reports (e.g., CTEDB, DC-CNN). Refer to Table~\ref{Table:rq2stats} for the test details. 

\looseness=-1
From our above investigation, we see consistent improvement in detecting textually dissimilar duplicate bug reports by the existing techniques, which might be explained as follows. First, the enrichment of bug reports with domain term explanations increases keyword overlap between duplicate reports, which might have benefited BM25's term-matching capability. Second, the enrichment also enhances the semantic representation of a bug report by adding contextual information, which might have helped like LDA+GloVe better capture the underlying meaning despite surface-level textual differences. Similarly, the deep learning models (SiameseCNN, DC-CNN, SBERT and CTEDB) can learn more robust representations from the augmented feature space. Finally, CUPID's hybrid architecture leverages its LLM component (ChatGPT) to better understand and process the enriched domain-specific context, while its traditional components benefit from the increased lexical overlap, resulting in improved detection of textually dissimilar duplicates. \vspace{-1mm}

\begin{boxH}
\textbf{Summary of RQ2: }
\looseness=-1
The enrichment of bug reports improved the performance of all seven baseline techniques in detecting textually dissimilar duplicates. While BM25 exhibit the maximum gains (e.g., 55.87\% for RecallRate@1), all techniques demonstrate significantly much higher improvements. These significant improvements across diverse techniques demonstrate the benefits of incorporating domain term explanations in duplicate bug report detection.
\end{boxH}

\section{Key Findings \& Insights}
\looseness=-1
\textbf{Software practitioners:}
Our findings 
deliver several actionable insights for software practitioners (e.g., developers, bug reporters).

\noindent
\textbf{(a) Improved bug report management:} Our idea of domain term extraction and explanation generation could have important implications. For example, it could be integrated into bug-tracking systems, providing real-time suggestions and context to bug reporters. According to existing work~\cite{zhang2017bug}, about 78\% of bug reports contain less than 100 words each, suggesting reporters' challenges in writing comprehensive bug reports. Our provided explanations can help them write better bug reports by explaining their domain terms and jargon.

\noindent
\textbf{(b) Enhanced Search:} Bug tracking systems (e.g., Bugzilla) contain up to 42\% duplicate bug reports~\cite{zou2018practitioners}, causing significant maintenance overhead. Bug reporters are not often aware that the same bug has been reported earlier, thanks to the text-based search engines of bug tracking systems. Our idea of detecting and explaining domain terms could enhance their search capabilities by delivering complementary information through explanations (e.g., synonyms and similar concepts). Our findings suggest that the explanations were effective in detecting textually dissimilar duplicate reports. Thus, it also has the potential to prevent reporters from submitting duplicate bug reports.

\noindent
\textbf{(c) Bug Report Comprehension:} According to an existing work~\cite{zhang2017bug}, short bug reports took 120 days extra to get resolved by the developers. The enrichment of bug reports with domain-term explanations can improve bug report comprehension, especially for complex technical issues that involve multiple domain-level concepts. This could benefit novice developers, who often struggle to contribute to open-source systems~\cite{xiao2022recommending}.

\noindent
\textbf{(d) Broader Applications in Software Engineering:} Our idea of domain term explanation can be extended beyond bug reports to other software artifacts like requirements documentation, code comments, and pull requests, where domain knowledge is equally crucial for understanding. \\
\noindent
\textbf{Researchers:}
Our work also encourages several research avenues.

\noindent
\textbf{(a) Extraction Optimization:} Future research could explore advanced techniques for domain term extraction by investigating contextual pattern mining, cross-artifact term relationships, and project-specific terminology evolution~\cite{chen2017understanding}. This is particularly crucial for understanding how domain terms manifest differently across various software artifacts. Further studies could systematically evaluate the quality and correctness of the generated explanations for domain terms. 

\noindent
\textbf{(b) Project Ecosystem Analysis:} Comparative research could examine how domain terminology varies between open-source and industrial projects, considering factors like project maturity, team distribution, and technical domain. This would help understand terminology patterns across different development contexts and improve documentation clarity and consistency~\cite{mao2017survey, robillard2017demand}. Such understanding can support software maintenance and evolution by reducing ambiguity and enhancing developer comprehension~\cite{steinmacher2016overcoming}.

\noindent
\textbf{(c) Impact on other Software Engineering tasks:} Researchers could investigate how domain-term-enriched documentation influences downstream development tasks. This includes measuring improvements in bug localization accuracy, prediction model performance, and issue resolution efficiency. 


\section{Related Work}\label{RelatedWork}

\textbf{Bug Report Enhancement:} Bug reports often suffer from incompleteness and a lack of clarity, prompting various enhancement approaches. Zhang et al.~\cite{zhang2017bug} developed an information retrieval approach that enhances bug reports with ranked sentences from historical data, addressing the challenge of short bug reports that delay resolution. Moran et al.~\cite{moran2018enhancing} developed FUSION for Android bug reports, focusing on reproduction steps. Wu et al.~\cite{wu2023intelligent} introduced CTEDB, combining term extraction with BERT-based duplicate bug report classification. However, their approach requires similar domain terminologies between reports. While these approaches make important contributions to bug report enhancement, they do not address the challenges of explaining domain terms prevalent in software bug reports. 

\looseness=-1
\textbf{Duplicate Bug Report Detection:} Duplicate bug report detection has evolved significantly, with various approaches addressing different aspects of the challenge. Traditional approaches like BM25F by Sun et al.~\cite{sun2011towards} relied on textual similarity and product information features. Isotani et al.~\cite{isotani2021duplicate} advanced the field with sentence embedding-based similarity, though limited to specific maintenance contexts. Chaparro et al.~\cite{chaparro2019reformulating} contributed query reformulation strategies by identifying software behaviour using discourse patterns which may not generalize. Recent work by Zheng et al.~\cite{zheng2024duplicate} proposed CorNER, a comprehensive approach combining NER and Random Forest algorithm to detect duplicate reports. However, they do not consider the aspect of textual dissimilarity between duplicate pairs. A recent empirical study by Jahan et al.~\cite{jahan2023towards} revealed that a significant proportion (19-23\%) of duplicate bug reports are textually dissimilar, where the current detection approaches struggle. While their work identifies this problem, our research takes a step further by directly addressing it through domain terminology enrichment. In our empirical study, we demonstrate how bug reports can be augmented with their domain-term explanations and how the semantic gaps between textually dissimilar duplicates can be bridged. 

\looseness=-1
Our work differentiates itself by addressing the challenge of textually dissimilar duplicate reports through report enrichment. Unlike previous approaches that depend heavily on textual similarity or require similar terminology between reports, we leverage comprehensive domain-specific vocabularies and transformer-based models to explain domain terms in a bug report. Our idea bridges the semantic gap between reports that describe the same issue using different technical terminology, addressing a critical limitation of existing duplicate detection methods. While previous works have extracted technical terms, our approach uniquely focuses on explaining these terms to enhance semantic understanding, particularly benefiting the detection of textually dissimilar duplicates that comprise a significant portion of duplicate reports (e.g., 23\%).

\section{Threats to Validity} \label{threats}
Threats to \textit{external} validity refer to the lack of generalizability in the findings~\cite{ferguson2004external}. One threat could stem from our selection of data sources. We select the API documentation, glossary and Stack Overflow tags for Java, which might not represent all relevant sources for software-specific terms or jargon. However, the underlying approach is not bound to any programming language and thus can be easily adapted to any other data sources or programming language.  

\textit{Construct} validity refers to the extent to which the experiment measures what it intends to measure~\cite{smith2005construct}. 
Inappropriate use of evaluation metrics could be a threat to construct validity. However, we chose our evaluation metrics --- Recall Rate @ k, Precision, Recall, F1 Score and AUC --- based on relevant literature~\cite{runeson2007detection,wang2008approach,sureka2010detecting,yang2012duplication, aggarwal2017detecting, alipour2013contextual, gopalan2014duplicate, tian2012improved, sun2010discriminative, klein2014new, he2020duplicate,budhiraja2018dwen, jahan2023towards}. Thus, threats to construct validity might be mitigated. 

Threats to \textit{internal} validity relate to experimental errors and subjective biases~\cite{christ2007experimental}. A source of threat could be the replication of the baseline techniques. We utilize the replication packages provided by the authors for LDA+GloVe~\cite{akilan2020fast}, SiameseCNN~\cite{deshmukh2017towards}, DC-CNN~\cite{he2020duplicate}, SBERT~\cite{reimers2019sentence}, and CUPID~\cite{zhang2023cupid}. For BM25, we use the replication package provided by Jahan et al.~\cite{jahan2023towards}. To our knowledge, no replication package is publicly available for CTEDB~\cite{wu2023intelligent}. We replicate it based on the methodology provided by the authors in the paper~\cite{wu2023intelligent}. For the replication of T5~\cite{raffel2020exploring}, we collected the pre-trained model from HuggingFace~\cite{huggingface_t5}. Furthermore, we followed the documentation closely for any customizations. Thus, threats to internal validity might be mitigated. The effectiveness of these models is dependent upon the quality and representativeness of the pre-training data. Issues such as biases present in the pre-trained models or their limited understanding of certain domain-specific terminology could impact the accuracy of explanations generated for bug reports.

\section{Conclusion and Future Work}
\looseness=-1
Automated detection of duplicate bug reports has advanced significantly over the years. However, research indicates that up to 23\% of duplicate bug reports are not textually similar. Additionally, around 78\% of bug reports in open-source projects are short (e.g., fewer than 100 words) and often include domain-specific terminology,
making it challenging to detect their duplicates. In this paper,
we conduct a large-scale empirical study to investigate the impact of domain terms and their explanations on the detection of duplicate bug reports. We analyze 92,854 bug reports from three open-source systems, replicate seven established baseline techniques for detecting duplicate bug reports, and answer two research questions. We find that enriching bug reports improves the performance of existing baselines significantly. Furthermore, the enrichment improved the performance across all detection techniques, with significant improvements in textually dissimilar duplicates. In the future, we plan on exploring enriching bug reports with multimedia content and code examples to enhance duplicate detection further.

\bibliographystyle{IEEEtran}
\bibliography{ref.bib} 

\end{document}